\documentstyle[11pt,newpasp,twoside,multicol,psfig]{article}
\def\edcomment#1{\iffalse\marginpar{\raggedright\sl#1\/}\else\relax\fi}
\reversemarginpar

\begin{document}
\title{X-ray Statistical Properties of the central cool component
in clusters of galaxies}

\author{Yasushi Ikebe}
\affil{Max-Planck-Institut f\"{u}r extraterrestrische physik,
           Postfach 1312, 85741 Garching, Germany}

\begin{abstract}
Central cool gas component that is often observed from
a well-relaxed cluster system
has long been interpreted as a consequence of ``Cooling Flow'' (CF),
radiative cooling followed by inflow of Intra-Cluster Medium (ICM).
However, recent XMM-Newton spectroscopy has shown 
no signatures of cooler gas phases below certain temperatures 
in typical CF clusters
(A1795, Tamura et al. 2001; A1835, Peterson et al. 2001).
This contradicts the conventional CF model
or at least requires a major revision of the model.
In order to investigate statistical properties of the central cool component,
we performed systematic analysis of {\it ASCA} data on 85 clusters.
We found that 1) temperature of the central cool component strongly
depends on the temperature of the main ICM,
2) the cool component is selectively found around a brightest cluster galaxy
(BCG) that coincide with the X-ray peak position,
and 3) the luminosity-temperature ($L$--$T$) relation of the cool component
shows nice agreement with the $L$--$T$ relation of the main ICM.
Together with the previous observational fact that,
in some of the ``CF'' clusters, the total gravitating mass is clustering
in two distinct spatial scales, a main cluster component
and a second small-scale system,
we conclude that the central cool component is associated with
the second small-scale self-gravitating system
that is immersed in the host cluster,
and the cool component temperature reflects the gravitational potential depth.
\end{abstract}

\section{Sample and ASCA analysis}

For the study we used brightest 85 clusters observed with {\it ASCA}
whose redshifts distribute from 0.004 to 0.201 with a median of 0.046.
From each GIS and SIS instrument, a spectrum over a central region
of $2'$ radius and one outside $2'$ radius were accumulated.
We fitted each set of {\it ASCA} spectra
with the two-temperature model (2T model),
in which a hot component is filling entire cluster region,
while in the central region a cool component is allowed to coexist
with the hot component forming two-phase plasma.
We then obtained X-ray luminosity in 0.1-2.4~keV band 
and temperature of cool and hot component,
$L_{\rm c}$, $T_{\rm c}$, $L_{\rm h}$, and $T_{\rm h}$.
Full results are given in Ikebe et al. (2001).
An example of the 2T model fit is shown in Fig.~1
for Abell~1795, which has one of the brightest central cool component.

\begin{figure}
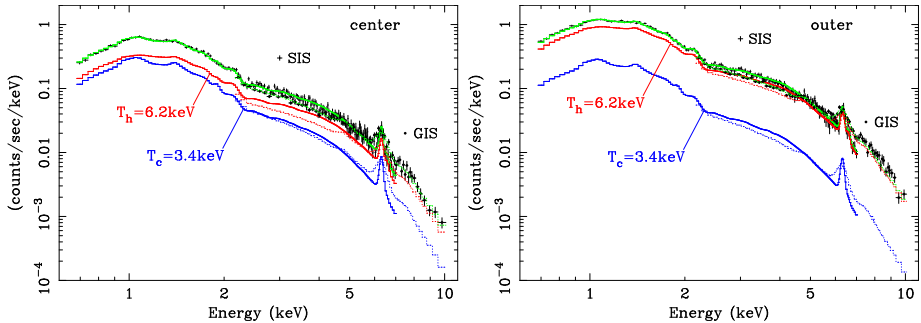

\centerline{
\psfig{file=ikebe1fig1.ps,width=6.0cm,angle=-90,clip=}
\psfig{file=ikebe1fig2.ps,width=6.0cm,angle=-90,clip=}}
\caption{
{\it ASCA} spectra of Abell~1795 extracted from the center (left)
and outer (right) regions.
Crosses and filled circles show SIS and GIS data, respectively.
Flux mixing effects due to the largely extended point-spread functions 
of X-Ray Telescope have been taken into account 
and the four spectra are fitted simultaneously with the 2T model.
The best-fit model for the SIS and GIS are illustrated
with solid and dotted lines respectively.}
\end{figure}

\section{Validity confirmation of the 2T model with XMM-Newton data}

The validity of the 2T model was tested with the latest X-ray data of
Abell~1795 taken with XMM-Newton.
The 11 annular spectra of EPIC-PN were simultaneously fitted with the 2T model,
where each annular spectrum is modeled with sum of two isothermal component,
and each component is assumed to have common temperature throughout all radii.
The fit is good and the best-fit two temperatures are derived to be 
$T_{\rm c}$=3.3~keV and $T_{\rm h}$=6.4~keV,
which are compared with the temperature profile obtained with the single-phase
model (Fig.~2).
The $T_{\rm c}$ and $T_{\rm h}$ nicely agree with the central coolest 
temperature and that of the isothermal component in the outer region, 
respectively.
Therefore, the 2T modeling is valid to the most advanced X-ray data 
and usable to characterize
the property of the central cool component and the surrounding ICM.

\begin{figure}
\centerline{\psfig{file=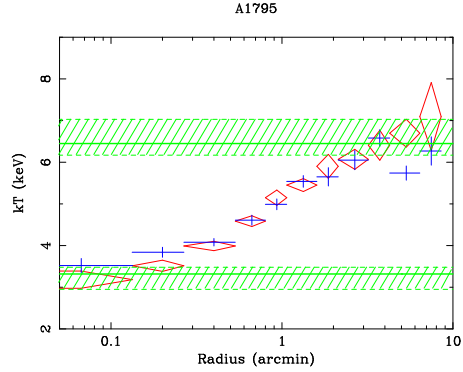,width=6.0cm,angle=-90,clip=}}
\caption{The temperature profile of Abell~1795.
The results from the single-phase model given in Tamura et al. (2001),
in which each annular spectrum is fitted with an isothermal plasma emission 
model, are reproduced with crosses (XMM/EPIC-MOS) and diamonds (PN).
The cool and hot component temperatures, $T_{\rm h}$ and $T_{\rm c}$,
derived with the 2T model are illustrated with crosshatched regions.
}
\end{figure}

\section{Results and Discussions}

Based on the results from the 2T model fitting to the {\it ASCA} data
of the 85 selected clusters, we investigated various correlations.
In the study, we classified the cluster sample into three groups,
which are Strong Cool Component (SCC),
X-ray Dominant (XD) and non-X-ray Dominant (nXD).
An SCC cluster is defined as one showing a very strong
(statistically very significant) cool component in the {\it ASCA} spectrum.
Among the rest of the clusters,
one in which the X-ray peak coincides with the Brightest Cluster Galaxy (BCG)
is classified as an XD cluster, while nXD is defined as one without
such a galaxy.
All the SCC clusters turned out to be XDs, too.

\begin{figure}
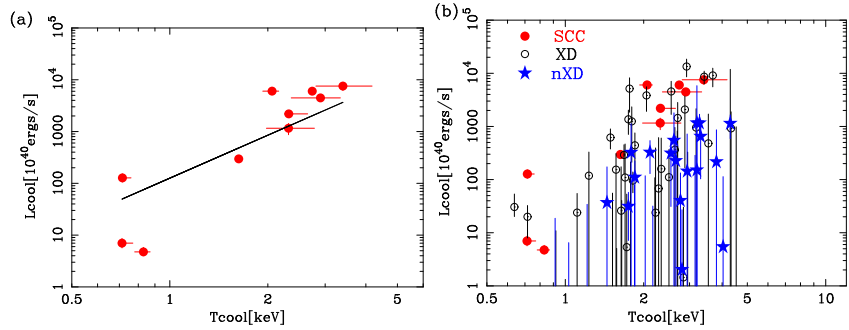

\centerline{
\psfig{file=ikebe3fig1.ps,width=5.5cm,angle=-90,clip=}
\psfig{file=ikebe3fig2.ps,width=5.5cm,angle=-90,clip=}}
\caption{
(a) The $L_{\rm c}$--$T_{\rm c}$ relation for the SCC clusters.
The line is the best-fit power-law function to the $L_{\rm h}$--$T_{\rm h}$
relation illustrated in Fig.~6.
(b) The $L_{\rm c}$--$T_{\rm c}$ relation for all the sample clusters.
Filled circles, open circles, and filled stars specify 
SCC, XD, and nXD clusters, respectively.}
\end{figure}

\begin{figure}
\centerline{\psfig{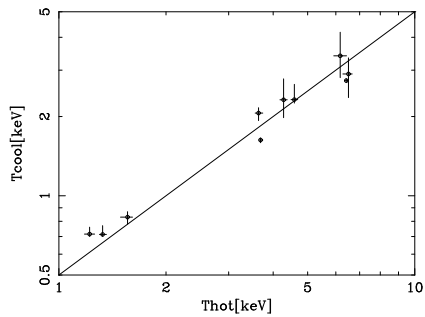}}
\caption{$T_{\rm c}$--$T_{\rm h}$ relation for the SCC clusters.
The line represents $T_{\rm c}$=$T_{\rm h}/2$.}
\end{figure}

For the SCC clusters, we determined the temperatures and emission measures
of each cool and hot component.
Figure~3(a) shows the cool component luminosity as a function of
the cool component temperature for the SCC clusters.
A positive correlation is clearly seen.
Figure~4 shows the correlation between the two temperatures
of the hot and cool component for the SCC clusters.
Surprisingly, the ratios between the two temperatures,
$T_{\rm c}/T_{\rm h}$, is virtually constant and
the relation, $T_{\rm c}$=$T_{\rm h}/2$ well represents the correlation.

For clusters that do not show very strong cool component, 
i.e. non-SCC clusters,
we fixed $T_{\rm c}$ at the half value of each mean temperature 
derived with a single-temperature model fitting,
according to the $T_{\rm c}$=$T_{\rm h}/2$ relation.
The luminosity of the cool component thus estimated for the non-SCC clusters
is illustrated in Fig.~3(b).
The results of the SCC clusters are also overlayed.
SCC and XD clusters exhibit systematically more luminous cool component 
than nXD clusters.
Actually, many of nXD clusters give only upper limit to $L_{\rm c}$.

We obtained the $L-T$ relation of the hot component,
which is illustrated in Fig.~5(a).
Unlike for the cool component, 
the three classes do not show any systematic difference.
It is then clearly shown that 
the X-ray characteristics of the XD and nXD clusters
are segregated mainly by the central cool component.
To compare with the $L-T$ relation of the cool component would
be very interesting.
In the Fig.~5(b), the $L_{\rm c}$--$T_{\rm c}$ relation for the SCC clusters
shown in Fig.~3 is overlayed on the $L_{\rm h}$--$T_{\rm h}$ relation.
A surprising agreement is clearly seen.
The best fit power-law function to the $L_{\rm h}$--$T_{\rm h}$ relation
is compared with the SCC $L_{\rm c}$--$T_{\rm c}$ relation in Fig.~3.
The similarity of the two $L$--$T$ relation suggests 
that both the cool and hot components
are related to individual gravitational bound objects.
Therefore, the cool component can be naturally interpreted as ICM filling
a self-gravitating system whose size is comparable to
a giant elliptical galaxy or a group of galaxies,
which is immersed in the host cluster.

\begin{figure}
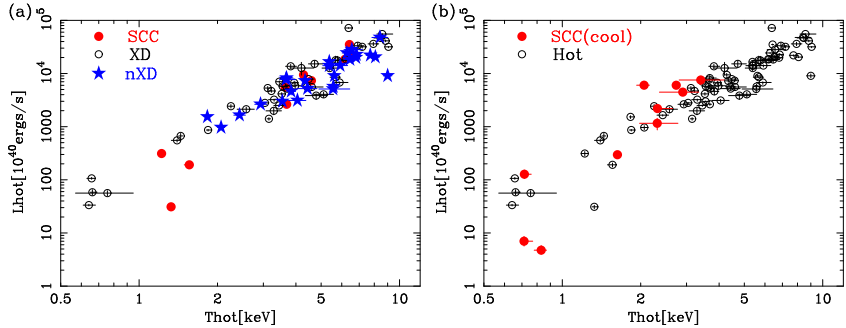

\centerline{
\psfig{file=ikebe5fig1.ps,width=5.5cm,angle=-90,clip=}
\psfig{file=ikebe5fig2.ps,width=5.5cm,angle=-90,clip=}}
\caption{(a) The $L_{\rm h}$--$T_{\rm h}$ relation for all the sample clusters.
Filled circles, open circles, and filled stars specify 
SCC, XD, and nXD clusters, respectively.
(b) The $L_{\rm c}$--$T_{\rm c}$ relation for the SCC clusters (filled circles)
is overlayed on the $L_{\rm h}$--$T_{\rm h}$ relation (open circles).}
\end{figure}

\end{document}